\documentclass[aps,prl,twocolumn,preprintnumbers,
amsmath,amssymb]{revtex4-1}
\usepackage{graphicx}
\usepackage{subfigure}
\usepackage{epsfig}
\usepackage{dcolumn}
\usepackage{bm}
\usepackage{ulem}
\usepackage{color}

\def\avg#1{\left\langle#1\right\rangle}

\def\be{\begin{equation}}       \def\ee{\end{equation}}
\def\bea{\begin{eqnarray}}      \def\eea{\end{eqnarray}}
\def\ba{\begin{array} }
\def\ea{\end{array} }
\def\bnum{\begin{enumerate} }
\def\enum{\end{enumerate}}

\def\nn{\nonumber}

\def\=>{\Rightarrow}
\def\>{\rightarrow}
\def\A{\uparrow}
\def\V{\downarrow}

\def\eye2{Fathbb{I}}

\def\Fig#1{Fig.~\ref{#1}}

\renewcommand{\>}{\rangle}

\begin{document}

\title{Realizing Majorana Zero Modes by Proximity Effect between Topological Insulators and $d$-wave High-Temperature Superconductors}

\author{Zi-Xiang Li}
\affiliation{Institute for Advanced Study, Tsinghua University, Beijing, 100084, China}
\author{Cheung Chan}
\affiliation{Institute for Advanced Study, Tsinghua University, Beijing, 100084, China}
\author{Hong Yao}
\affiliation{Institute for Advanced Study, Tsinghua University, Beijing, 100084, China}

\begin{abstract}
We theoretically study superconducting proximity effect between a topological insulator (TI) and a high-temperature $d$-wave superconductor (dSC). When the TI-dSC heterostructure violates $90^\circ$-rotation and certain reflection symmetries, we show that a sizable $s$-wave pairing, coexisting with a $d$-wave one, emerges in the proximity-induced superconductivity in the TI's top surface states. Weak disorder further suppresses $d$-wave pairing but not $s$-wave one in the TI's surface states. More importantly, the pairing gap in surface states is found to be nodeless and nearly-isotropic when the Fermi pocket of surface states is relatively small. Our theoretical results qualitatively explain recent experimental evidences of a nearly-isotropic pairing gap on surface states of Bi$_2$Se$_3$  induced by proximity to high-$T_c$ cuprate Bi$_2$Ca$_2$Cu$_2$O$_{8+\delta}$. We also demonstrate convincing evidences of Majorana zero modes in a magnetic $hc/2e$ vortex core, which may be detectable in future experiments.
\end{abstract}
\date{\today}

\maketitle
{\bf Introduction:} Searching for Majorana zero modes in condensed matter systems has attracted great attentions in the past few years\cite{Jason-12,Beenakker-13}. Because they obey non-Abelian statistics\cite{Moore-Read,Ivanov-01}, it is widely believed that they could play a key role in realizing topological quantum computation\cite{Kiteav-03,Nayak-Simon-Stern-Freedman-Sarma-08}. Since Fu and Kane's original proposal of employing proximity effect between the surface states of a 3D topological insulator (TI)\cite{Qi-Zhang,Hasan-Kane} and a $s$-wave superconductor (sSC)\cite{Fu-Kane-09}, a number of alternative approaches has been proposed to realize such non-Abelian excitations \cite{Sau-10,Lutchyn-10,Oreg-10,Jason-10,Qi-Hughes-10,Ashvin-11} and much progress has been made experimentally\cite{Mourik-12,LiLv,Das-12,Rokhinson-12,Deng-12,Chang-13,Finck-13}. By proximity to a sSC, an effective $p+ip$-like pairing can be induced on the TI's spin-momentum-locked surface states; consequently a $hc/2e$ magnetic vortex hosts a Majorana zero mode\cite{Kopnin-Salomaa-91,Read-Green-01,Tewari-Sarma-Lee-08}. Exciting experimental progress has been reported in realizing such proximity effect between TI and convectional sSC\cite{JinfengJia-12,JinfengJia-13}. However, it is challenging to identify Majorana zero modes in such systems mainly because of the small pairing gap in the TI's surface states\cite{JinfengJia-12}.
With such small gap, the energy separation between Majorana zero mode and excited states in a vortex core is estimated to be about $0.001$meV, which is hardly resolvable by instruments like a state-of-art STM.

A natural approach to remedy the shortcoming discussed above is to employ a high $T_c$ cuprate superconductor\cite{Kivelson-Stripe}, which have much larger gap compared with conventional superconductors, in proximity to TIs. Indeed, efforts along this direction have been made\cite{Zareapour-12,ShuyunZhou-13}. Especially, high quality Bi$_2$Se$_3$ thin films has been successfully grown on fresh surface of the cuprate superconductor Bi$_2$Ca$_2$Cu$_2$O$_{8+\delta}$ (Bi2212) by MBE\cite {ShuyunZhou-13}, for which ARPES measurements show evidences of a nearly-isotropic proximity-induced gap of about 10meV in the TI's top surface states, as shown in Fig. 1(a). Such heterostructure with nodeless pairing gap on the TI's surface states may provide a promising arena for realizing Majorana zero modes and possibly implementing quantum computations. Nonetheless, there are two important questions which need to addressed. Firstly, since the driving force of such proximity is a nodal $d$-wave superconductor (dSC), we would expect that the proximity-induced gap on the TI's Fermi surfaces carry nodes (see {\it e.g.} Ref.\cite{Nagaosa-10,Annica}); is it possible to have a nodeless and nearly-isotropic gap induced on the TI's surface states by proximity to a dSC? Secondly, even when a nodeless gap is induced on the TI's top surface states, can such heterostructure harbor robust and localized Majorana zero modes in its magnetic vortex core considering that the dSC has nodal quasiparticle excitations? In this paper, we theoretically answer these questions.

The superconducting order parameter in the $d$-wave cuprate is {\it odd} under either $90^\circ$-rotation or reflection along the $x'z$ plane ($\hat x'\propto\hat x\pm\hat y$). When a TI is in proximity to the $d$-wave cuprate, $s$-wave pairing in the TI must be zero and the pairing gap in the TI's surface states must carry nodes if the TI respects either $90^\circ$-rotation or $x'z$-reflection. Fortunately, the TI Bi$_{2}$Se$_{3}$ is a hexagonal system not respecting the $90^\circ$-rotation. Moreover, a relative orientation between the TI and the cuprate can be chosen such that it maximally violates the $x'z$ reflection symmetry, as observed by STM measurements\cite{ShuyunZhou-13}. This mismatch between the two lattice symmetries makes it possible that a finite $s$-wave pairing coexists with a $d$-wave one in the TI. Indeed, through self-consistent mean-field calculations, we find that a sizable $s$-wave pairing emerges in the TI because the heterostructure has neither $90^\circ$-rotation nor $x'z$-reflection. Moreover, since the Fermi pockets of the TI's surface states is relatively small, gap anisotropy generated by the $d$-wave pairing component is negligible and the pairing gap is dominated by the $s$-wave component; consequently, a nodeless and nearly-isotropic pairing gap in the TI's top surface states can be realized. Moreover, we find convincing evidences of Majorana zero mode localized in a magnetic vortex core, which may be detected by future STM measurements.

\begin{figure}[t]
\includegraphics[width=3.0in]{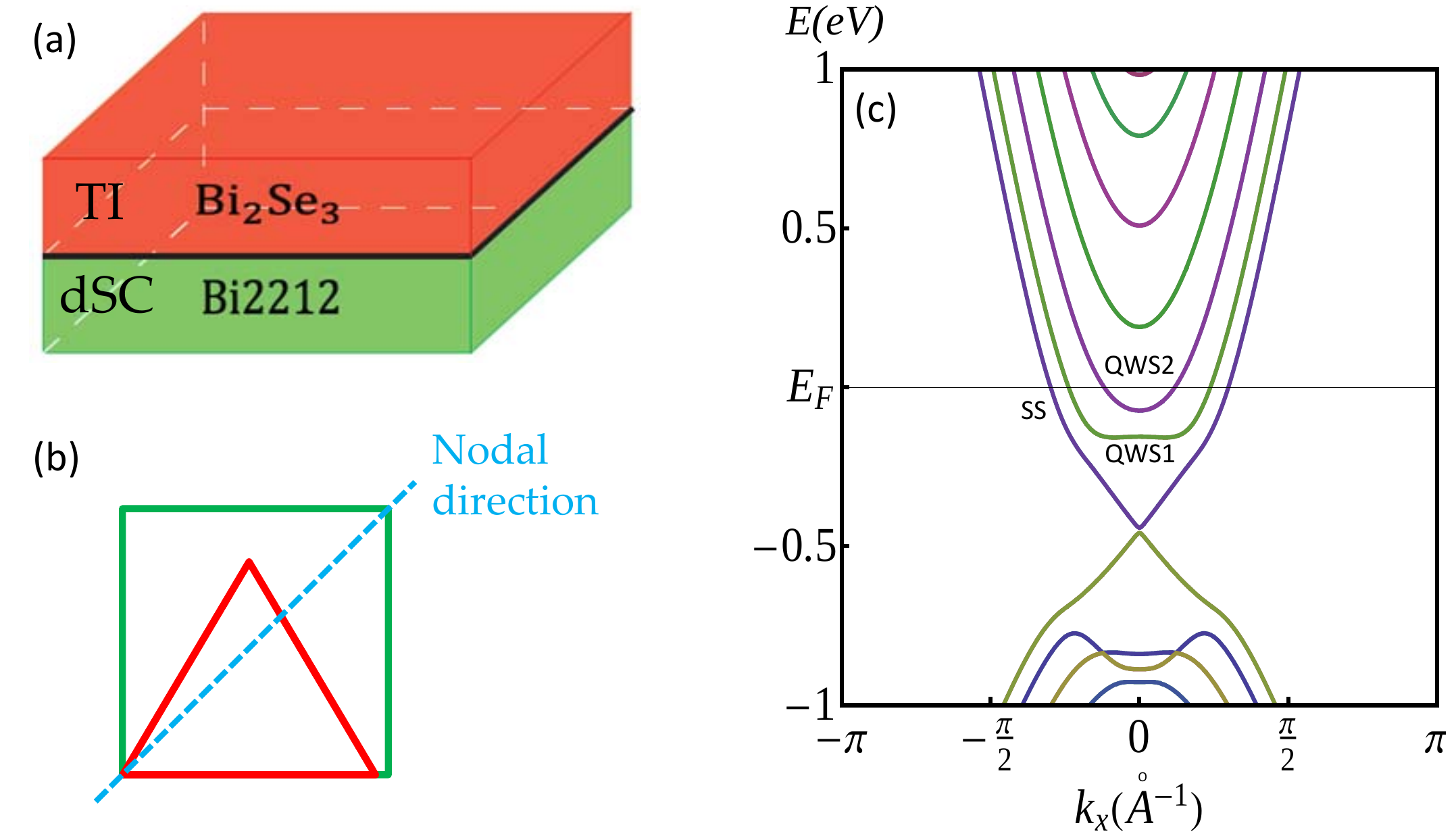}
\caption{(a) Schematic representation of a TI-dSC heterostructure. (b) The relative lattice orientation between the TI Bi$_2$Se$_3$ (triangular lattice) and the dSC Bi2212 (square lattice). The heterostructure violates both  $90^\circ$-rotation and reflection along nodal directions.  (c) The TI's band structure with its Fermi level tuned to across top surface states (SSs) as well as two quantum well states (QWSs).}
\label{fig:lattice}
\end{figure}

{\bf Model:} We consider a TI-dSC heterostructure with Hamiltonian
$H = H^\textrm{TI}_0+H^\textrm{TI}_1 + H^\textrm{dSC}_0+H^\textrm{dSC}_1 + H'$,
where $H^\textrm{TI}=H^\textrm{TI}_0+H^\textrm{TI}_1$ describes a TI, $H^\textrm{dSC}=H^\textrm{dSC}_0+H^\textrm{dSC}_1$ a $d$-wave superconductor, and $H'$ represents couplings between them. Noninteracting parts of $H$ are given by:
\bea
&&H^\textrm{TI}_0 =\sum_{\vec k} c^\dag_{\vec k} \Big[
E(\vec k)\tau_{0}\sigma_{0}+ M(\vec k)\tau_{z}\sigma_{0} + A_{1} \sin k_{z} \tau_{x}\sigma_{z}  \nn\\
&&~~~~~~~~~~~~~ + A_{2}\sin k_{x}\tau_{x}\sigma_{x} + A_2\sin k_{y}\tau_{x}\sigma_{y})\Big]c_{\vec k}, \\
&&H^\textrm{dSC}_0 = \sum_{\vec k\sigma} \xi(k) d^{\dag}_{\vec k\sigma}d_{\vec k\sigma},\\
&&H' = \sum_{\vec k\sigma}\left[ t'_{1} d^\dag_{\vec k\sigma}c_{\vec k,1\sigma} + t'_{2} d^{\dag}_{\vec k\sigma}c_{\vec k,2\sigma} + H.c.\right],
\eea
where $c^\dag_{\vec k}=(c^\dag_{\vec k1\A},c^\dag_{\vec k1\V},c^\dag_{\vec k2\A},c^\dag_{\vec k2\V})$, and $\tau_\alpha$ and $\sigma_\alpha$ are identity or Pauli matrices with orbital indices $\tau=1,2$ and spin indices $\sigma=\A,\V$, respectively. Here $c_{\vec k\tau\sigma}$ and $d_{\vec k\sigma}$ annihilate an electron in the TI and in the cuprate, respectively.
For simplicity, we consider a TI on a cubic lattice with a Hamiltonian violating both $90^\circ$-rotation and $x'z$ reflection symmetries, which mimics the experimental situation\cite{ShuyunZhou-13,domain-wall}.
We assume $ M(\vec k) = M + 2t_x+ 2t_y + 2t_z - 2t_{x} \cos k_{x} - 2t_{y}\cos k_{y} - 2t_{z}\cos k_{z}  $ and $ E(k) = 2\tilde t_{x} + 2\tilde t_{y} + 2\tilde t_{z} - 2\tilde t_{x} \cos k_{x}  - 2\tilde t_{y} \cos k_{y}  - 2\tilde t_{y} \cos k_{z} -\mu $ with $M=-0.4$eV, $t_{x} = 0.5$eV, $t_{y}= 0.25$eV, $t_{z} = 0.4$eV, $\tilde t_{x}=0.1$eV, $\tilde t_y=0.05$eV, $\tilde t_{z}=0.1$eV, $A_{1}=0.4$eV, and $A_{2}=0.4$eV\cite{HaijunZhang-09}, with which $H_0^\textrm{TI}$ describes a TI with nontrivial surface states, as shown in \Fig{fig:lattice}(b). We emphasis that this effective TI Hamiltonian with $t_x\neq t_y$ and $\tilde t_x \neq \tilde t_y$ breaks both $90^\circ$-rotation and $x'z$ reflection symmetries. The lack of such symmetries in the heterostructure plays a key role in inducing a finite $s$-wave pairing in the TI's surface states.

\begin{figure}[t]
\centering
\includegraphics[width=3.3in]{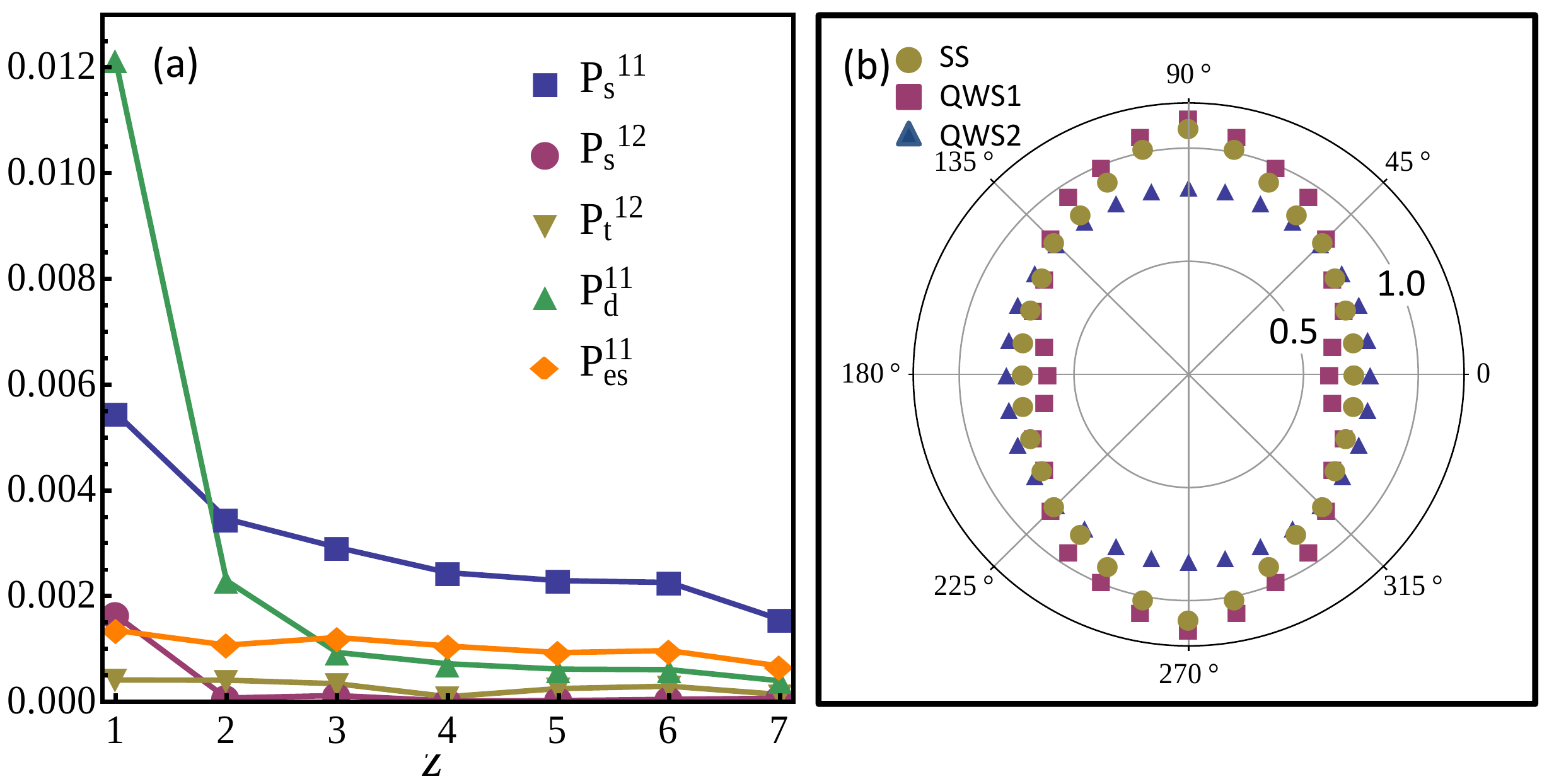}
\caption{(a) Various pairing amplitudes as a function of $z$, the layer index of the TI. Here $z=1$ labels the layer close to the interface and $z=7$ is the top layer. (b) Pairing gaps induced on the TI's top SSs, QWS1, and QWS2 as a function of angle. All of them are nodeless and nearly-isotropic.
}
\label{fig:gap}
\end{figure}

For the $d$-wave cuprate, its effective band dispersion is given by $\xi(k) = -2 t_1(\cos k_{x}+\cos k_{y}) -4 t_2\cos k_{x}\cos k_{y}-\mu'$ with $t_1= 0.25$eV, $t_2= -0.05$eV; $\mu'=-0.18$eV at $15\%$ hole doping. Vertical hoppings between the TI and the cuprate are $t'_1=t'_2=0.5$eV. As for $H^\textrm{dSC}_1$, we assume effective attractive interactions which favor singlet $d$-wave pairing:
\bea
H^\textrm{dSC}_1 = \sum_{i,\delta=\hat x,\hat y} J(\vec S_i\cdot \vec S_{i+\delta} - \frac14 n_in_{i+\delta}),  ~~\nn
\eea
where $J$ is the interaction strength and then study superconductivity through a self-consistent mean-field treatment\cite{footnote1}. We set $J=0.4$eV so that an antinodal gap of about 50meV is obtained in the mean-field theory of isolated cuprate itself. We further consider effective on-site interactions in the TI:
\bea
H^\textrm{TI}_1= U\sum_{i\tau} c^{\dagger}_{i\tau\uparrow}c_{i\tau\uparrow} c^{\dagger}_{i\tau \downarrow}c_{i\tau\downarrow}
 \!+\! V\sum_{i\sigma\sigma'}  c^{\dagger}_{i1\sigma}c_{i1\sigma} c^{\dagger}_{i2\sigma'}c_{i2\sigma'},~~~\nn
\eea
where $U$ ($V$) are intra-orbital (inter-orbital) interactions that may be repulsive or attractive. Effective attractions in the TI can be induced by electron-phonon couplings; they should be weak and do not directly generate superconductivity at temperatures experimentally reached. Note that nature of weak interactions in the TI does not qualitatively change our main results as discussed below.

We employ self-consistent mean-field treatment to deal with interactions in the dSC and the TI. After mean-field decomposition, we obtain:
\bea
 &&H^\textrm{dSC}_{1,\textrm{MF}}\! =\! \sum_{\vec kz} \left[\Delta_x(z)\cos k_{x}\!+\!\Delta_y(z)\cos k_y \right]d_{\vec k z\uparrow}^{\dagger}d_{-\vec kz\downarrow}^{\dagger} \!+\! H.c.,\nn\\
 &&H^\textrm{TI}_{1,\textrm{MF}}\! = \!\sum_{\vec kz}\Big[\Delta_{s}^{11}(z)c_{\vec k z1\uparrow}^{\dagger}c_{-\vec kz1\downarrow}^{\dagger}+\Delta_{s}^{22}(z)c_{\vec kz2\uparrow}^{\dagger}c_{-\vec kz2\downarrow}^{\dagger}\nn\\
 &&+ \Delta_{s}^{12}(z)i\sigma^y_{ss'}c_{\vec kz1s}^{\dagger}c_{-\vec kz2s'}^{\dagger}\!+\! \Delta_{t,0}^{12}(z)\sigma^x_{ss'}c_{\vec kz1s}^{\dagger}c_{-\vec k z2s'}^{\dagger} \!+\! H.c.\Big],\nn
\eea
where pairing potentials are given by $\Delta_\delta(z)=J \langle d_{\vec k z\V}d_{-\vec k z\A}\rangle\cos k_\delta$ ($\delta$=$x,y$), $\Delta^{11}_s(z)=UP^{11}_s(z)$, $\Delta^{22}_s(z)=UP^{22}_s(z)$,
$\Delta^{12}_s(z)=VP^{12}_{s}(z)$ and $\Delta^{12}_{t,0}(z) = VP_{t,0}^{12}(z)$, which shall be solved self-consistently. Here $P$ label various pairing amplitudes in the TI: 
$P^{\tau\tau'}_{s}= (i\sigma^y)_{\sigma\sigma'}\avg {c_{i\tau\sigma}c_{i\tau'\sigma'}}$ and $P^{\tau\tau'}_{t,0} =\sigma^x_{\sigma\sigma'}\avg {c_{i\tau\sigma}c_{i\tau'\sigma'}}$ are on-site pairing amplitudes while $P^{\tau\tau'}_{d}=$$[(i\sigma^y)_{\sigma\sigma'}\avg{c_{i\tau\sigma}c_{i+\hat x,\tau'\sigma'} -c_{i,\alpha\sigma}c_{i+\hat y,\beta\sigma'} }]/2$ and $P^{\tau\tau'}_{es} = [(i\sigma^y)_{\sigma\sigma'}\avg{c_{i\tau\sigma}c_{i+\hat x,\tau'\sigma'} +c_{i,\alpha\sigma}c_{i+\hat y,\beta\sigma'} }]/2$ are pairing amplitudes on nearest neighbor bonds. Note that the pairing potentials on the nearest-neighbor bonds in the TI are zero even though the pairing amplitudes on corresponding bonds can be nonzero.

{\bf Proximity effect:} A Bi$_2$Se$_3$ thin film with nominally 7 quintuple layers (QL) is grown successfully on the top of a bulk cuprate Bi2212\cite{ShuyunZhou-13}. For the TI thin film with 7QL, its surface states are  localized on the TI's top surface while its interface states are localized at the interface of the heterostructure; consequently they are nearly decoupled and remain gapless\cite{decoupled}.
Besides top surface states (SSs) and interface states (ISs), a TI thin film has also quantum well states (QWSs). If the TI's Fermi level crossed the top SSs but not QWSs, we expect induced superconductivity on the TI's top surface to be exponentially weak. In other words, to induced appreciable superconductivity in the TI's top surface, its Fermi level needs to cross some QWSs, which can help propagating superconducting correlations to the TI's top surface. We set $\mu=0.55$eV so that the TI's Fermi level crosses top SSs as well as two bands of QWSs, as shown in \Fig{fig:lattice}(b).  Even though a bulk cuprate is used in experiments, we find that increasing the number of layers of cuprates larger than a few has negligible effect to the proximity phenomena induced in the TI. Hereafter, we consider a heterostructure consisting of a TI with 7 layers and a $d$-wave superconductor with 5 layers.

We solve self-consistent mean-field equations numerically to obtain the pairing gap on TI's SSs and QWSs by assuming $U= -0.25$eV and $V= -0.1$eV\cite{footnote2}. It's worth mentioning that when turning off the coupling between the TI and the cuprate, namely setting $H'=0$, the weak attractive interactions themselves in the TI induces negligibly-small pairing amplitudes on the TI's top surface, which are about 1000 times smaller than the induced one by proximity effect. This indicates that the pairing in the TI is mainly formed by proximity to the dSC other than by its weak attractive interactions.

With the interactions in the TI and dSC taking values above, we obtain various pairing amplitudes in the TI as a function of layer index $z$, as shown as \Fig{fig:gap}(a). Close to the TI-dSC interface, the $d$-wave pairing amplitude on nearest-neighbor bonds is stronger than the extended $s$-wave one, as expected. However, the further away from the interface, the smaller the ratio between the $d$-wave and extended $s$-wave ones, as shown in \Fig{fig:gap}(a). On the top layer, namely $z=7$, the extended $s$-wave component is comparable to the $d$-wave one; moreover, the on-site $s$-wave pairing amplitude becomes the most dominant one on the top layer. The appearance of finite $s$-wave pairing amplitudes in the TI is attributed to the breaking of both $90^\circ$-rotation and $x'z$-reflection symmetries in the heterostructure. When we set $t_x = t_y $ and $\tilde t_x = \tilde t_y$, namely imposing the symmetries by hand, $s$-wave pairing amplitudes in the TI vanish, as expected.

The pairing gap in the TI's top SSs and two bands of QWSs along different directions are plotted in \Fig{fig:gap}(b). It is clear that these pairing gaps are nodeless, which are dramatically different from the nodal gap in the $d$-wave cuprates. Naively, for pairings with both $s$-wave and $d$-wave components, a dominant $s$-wave one is needed to produce a nodeless pairing gap. However, gap anisotropy produced by the $d$-wave pairing $\Delta (\cos k_x-\cos k_y)$ sensitively depends on where the Fermi surface is. When the Fermi surface is small and close to the $\Gamma$ point, the gap anisotropy generated by it is enormously reduced by a factor of $(1-\cos k_F)$, where $k_F$ labels a characteristic radius of the Fermi surface. Indeed, the Fermi pockets of the TI's SSs and QWSs are extremely small; $k_F$ is about ${\pi}/{4}$ in our effective model and it is even smaller in real materials. As a result, the pairing gap on the TI's Fermi surfaces can be nodeless even when there were an appreciable $d$-wave pairing component. Our results of a nodeless and nearly-isotropic pairing gap in the TI's top surface states is consistent with what were observed by ARPES experiments\cite{ShuyunZhou-13}.

\begin{figure}[t]
\centering
\includegraphics[width=3.3in]{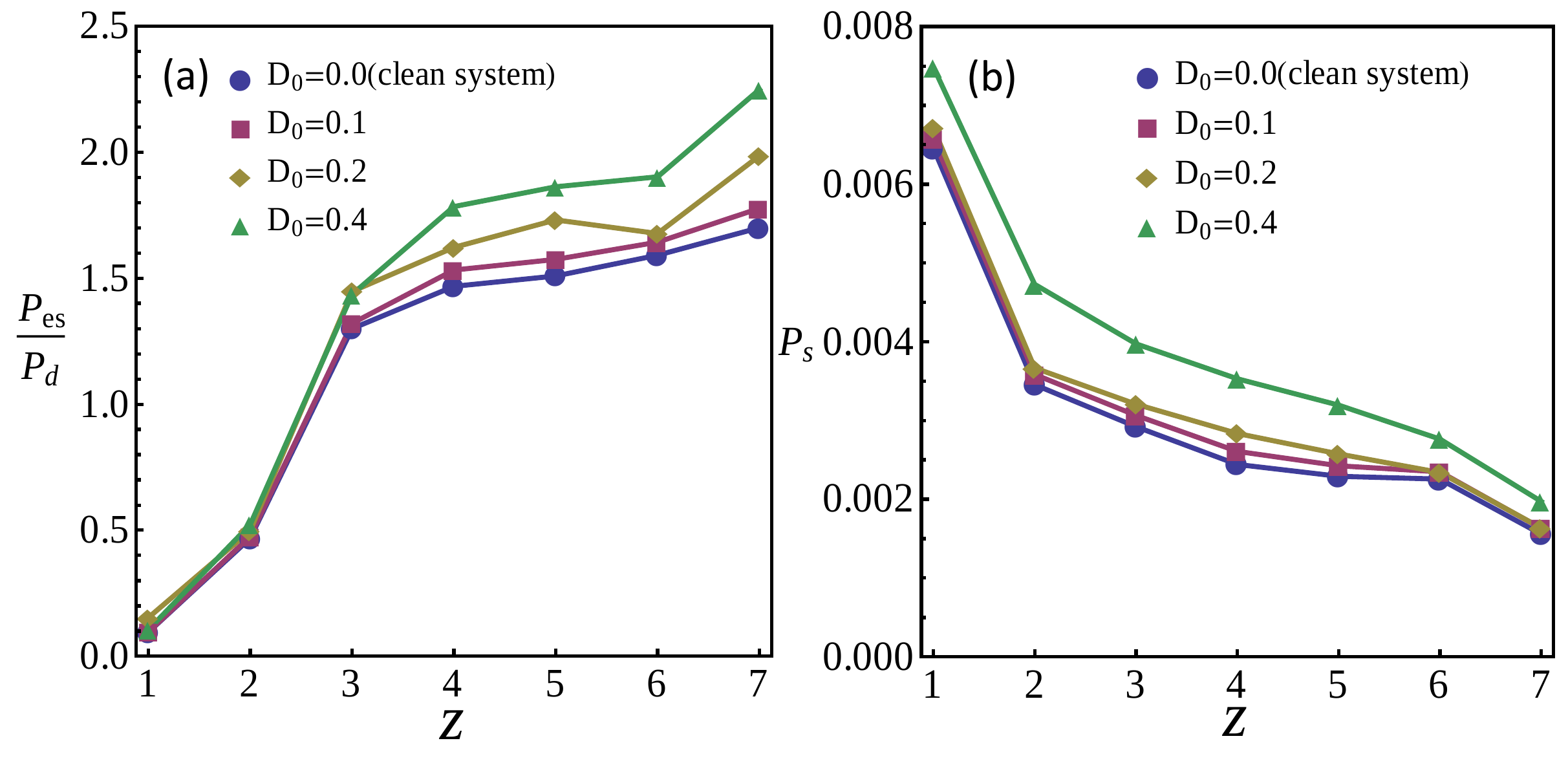}
\caption{(a) The ratio between the extended $s$-wave pairing and the $d$-wave pairing amplitudes and (b) The on-site $s$-wave pairing amplitude, for different disorder strengths.}
\label{fig:disorder}
\end{figure}

{\bf Effect of disorder:} It has been long known that disorder can enhance isotropy of pairing gap. We discuss the effect of disorder on the proximity-induced pairing gap in the TI. For simplicity, we consider a random chemical potential in the TI which obeys uniform distribution:
$\mu(\vec r_i) = \mu_{0} + \delta\mu(\vec r_i)$,
where $\mu_{0}$ is the chemical potential in the clean TI and $\delta\mu(\vec r_i)$ is the spatial random part with uniform distribution in the range of $[-D_0,D_0]$. Here $D_0$ characterizes the strength of disorder in the TI. With disorder, the hybrid system does not respect translational symmetry in the $xy$ plane; consequently it is quite computationally costly to solve the self-consistent mean-field equations in presence of disorder. Instead, we use the BdG Hamiltonian with pairing potentials obtained self-consistently in the clean limit and then study the influence of disorder to the pairing in the TI.

We expect that the anisotropic part of the pairing in the TI should be suppressed by disorder, while the on-site isotropic pairing is robust against disorder, according to Anderson's theorem\cite{Anderson}. This is indeed what we find numerically. When the disorder strength $D_0$ increases, the ratio between the extended $s$-wave pairing amplitude and the $d$-wave one increases\cite{Kivelson}, as shown in \Fig{fig:disorder}(a); the on-site $s$-wave pairing amplitude in the TI has little change, as shown in \Fig{fig:disorder}(b). Consequently, we have shown that disorder plays a role in making pairing gap more isotropic around the Fermi surfaces of the TI.

\begin{figure}
\centering
\includegraphics[width=3.2in]{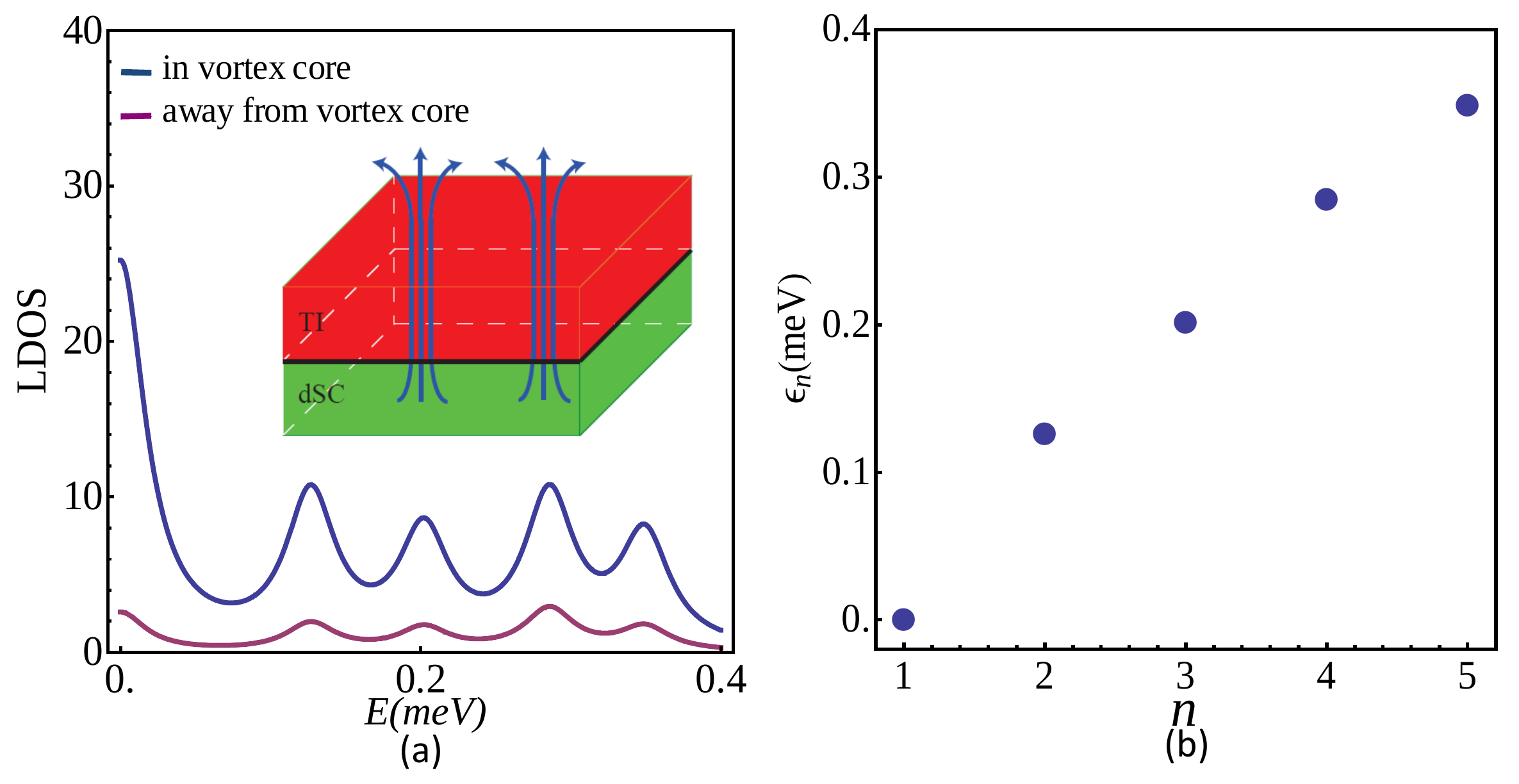}
\caption{(a) LDOS at and away from vortex centers. Two magnetic vortices are shown in the insert. (b) Energy eigenvalues in the presence of two magnetic vortices. A complex zero mode, which consists of two Majorana zero modes, appears and is separated from the lowest excited state by about 0.1meV. }
\label{fig:vortex}
\end{figure}

{\bf Majorana zero modes:} We have shown that a nodeless pairing gap can be induced in the TI's top SSs as well as QWSs by proximity to a dSC. Moreover, with disorder the gap can become more isotropic. It is then natural to ask whether such a TI-dSC heterostructure can support one Majorana zero mode in a magnetic vortex core. It is now well-known that a TI-sSC heterostructure can support a robust and localized Majorana zero mode in a magnetic $hc/2e$ vortex. Even though a nodeless pairing gap opens in the TI's surface states for both heterostructures, there is an important difference between them: the dSC in the substrate is nodal while the sSC is fully gapped. It is not {\it a priori} clear  whether a localized Majorana zero mode can exist in the presence of low-lying gapless quasiparticle excitations in the substrate.

To investigate this issue, we study the TI-dSC hybrid system in the presence of two well-separated vortices as shown in \Fig{fig:vortex}(a). The vortices are added by imposing winding phases in superconducting order parameters: $\Delta(\vec r_1,\vec r_2)=\Delta_0(\vec r_1,\vec r_2)\to  \Delta_{0}(\vec r_1,\vec r_2)e^{i \theta(\vec r_{1},\vec r_{2}) }$, where $\Delta_0(\vec r_1,\vec r_2)$ are obtained through the self-consistent mean-field calculations without vortices, $e^{i \theta(\vec r_{1},\vec r_{2}) }\equiv \frac{e^{i\theta(\vec r_{1})} + e^{i\theta(\vec r_{2})}}{|e^{i\theta(\vec r_{1})} + e^{i\theta(\vec r_{2})}|}$, and  $\vec r_{1,2}$ label two sites of a nearest-neighbor bond in the cuprate. Since we use periodic boundary conditions along $x$ and $y$ directions, a special gauge transformation \cite{Franz-00} is employed to ensure continuous superconducting order parameters across the boundaries. We take a 150$\times$30 lattice and insert two magnetic $hc/2e$ vortices separated from each other by 70 lattice constants along $x$ direction, as shown in \Fig{fig:vortex}(a). With these two vortices, we find that one complex zero-energy mode appears, with a sizable energy-separation from excited states, as seen in \Fig{fig:vortex}(b). The wave function of this zero-energy mode is localized around the two vortex cores and also localized near the TI's top surface, which is spatially separated from the gapless modes in the substrate of the dSC. This zero-energy complex mode is actually a pair of Majorana zero modes on the TI's top surface, each of which is localized in a vortex core. Note that the interface does not support well-defined Majorana zero modes because of the gapless excitations in the interface and in the dSC substrate.

The appearance of a robust and well-localized Majorana zero mode in the vortex core in the TI-dSC heterostructure  is very encouraging. To detect it experimentally, one can measure local density of states (LDOS) $\rho(\vec r,E)$ on the TI's top surface by STM according to $\rho(\vec r,E)\propto \frac{dI}{dV}(\vec r,E)$, where $\frac{dI}{dV}(\vec r,E)$ is differential conductance. We numerically compute the LDOS, which are contributed from top three layers of the TI. The LDOS at two typical positions, one at vortex core and the other away from vortex centers by 35 lattice constants, are plotted in \Fig{fig:vortex}(a). We see a dominant zero-bias peak in the LDOS at the vortex center, which is well-separated from the higher-energy peaks by about $0.1$meV; consequently this zero-bias peak may be resolved within the accuracy of a state-of-art STM. Note that this energy separation is much larger than that expected in a TI-sSC heterostructure\cite{JinfengJia-12}. When we move to a position 35 lattice constants away from vortex centers, the weight of zero-bias peak in its LDOS is much weaker, since the Majorana zero mode is localized in vortex cores. The localization of Majorana zero modes is a manifestation of the nodeless pairing gap in the TI's top SSs and QWSs.

{\bf Concluding remarks:} We have demonstrated that under certain lattice symmetry requirements a nodeless or even nearly-isotropic pairing gap can be achieved on TI's top SSs and QWSs by proximity to a dSC. Our results convincingly explain the seemingly surprising observations of nearly-isotropic pairing gap in the TI's top SSs. We also demonstrate that such a TI-dSC heterostructure can support Majorana zero modes. While our theoretical results show similar gap sizes in top SSs and QWSs, the ARPES experiment\cite{ShuyunZhou-13} reports that the pairing gap on  QWSs is significantly smaller than the one in SSs, which is still puzzling since QWSs are expected to have stronger superconducting proximity to the dSC than top SSs. Further experimental studies are desired to investigate such a TI-dSC system, which we believe provides a promising arena for realizing robust Majorana zero modes in 2D.

{\it Acknowledgement}: We sincerely thank Xi Chen, Liang Fu, Eun-Ah Kim, Steve Kivelson, K. T. Law, Dung-Hai Lee, and Shuyun Zhou for helpful discussions. This work is supported in part by the Thousand-Young-Talent Program of China.

{\it Note added}: During the preparation of our manuscript, we notice an interesting theoretical paper on a related but different topic which studies the proximity between a TI and a conventional sSC\cite{Kim-14}.


\begin{thebibliography}{10}
\bibitem{domain-wall} For the TI-dSC heterostructure studied in Ref. \cite{ShuyunZhou-13}, two TI domains with different lattice orientaions are found by STM meansurements. The lattice orientation of one TI domain is mapped to the other by reflection along $x'z$ plane. As a consequence, the $s$-wave pairing order parameter in one TI domain induced by proximity to a dSC is opposite to the one in the other TI domain. Then, along domain wall, there are gapless helical Majorana modes which might be detectable by STM experiments.
\bibitem{footnote1} If we treat $H^\textrm{dSC}_1$ as $\sum_{\vec k}\big[\Delta (\cos k_x-\cos k_y)d^\dag_{\vec k\A}d^\dag_{- \vec k \V}+H.c.\big]$ [namely treat $H^\textrm{dSC}$ as a Bogoliubov-de Gennes (BdG) Hamiltonian] and then study the proximity effect from it, nodes in the pairing gap in the TI's surface states must appear along the diagonal directions $\vec k_x=\pm \vec k_y$ because the paring potential along these directions vanishes even in the cuprate itself, not mentioning in the TI's surface states. In other words, a simple BdG Hamiltonian for the dSC is not sufficient for discussing the proximity effect in such hybrid TI-dSC system; we have to directly consider interactions in the cuprate and study superconductivity through a self-consistent mean-field treatment.
\bibitem{footnote2} If we consider weak repulsive interactions instead of weak attractive ones in the TI, a fully gapped Fermi surfaces in the TI's top SSs and QWSs is also obtained, even though the gap size is somewhat smaller than that with attractive interactions.
\end{thebibliography}
\end{document}